# Radiation pressure and the distribution of electromagnetic force in dielectric media (II)

**Armis R. Zakharian, Masud Mansuripur, and Jerome V. Moloney**

Optical Sciences Center, The University of Arizona, Tucson, Arizona 85721
*masud@optics.arizona.edu*

**Abstract**: Using the Finite-Difference-Time-Domain (FDTD) method, we compute the electromagnetic field distribution in and around dielectric media of various shapes and optical properties. With the aid of the constitutive relations, we proceed to compute the bound charge and bound current densities, then employ the Lorentz law of force to determine the distribution of force density within the regions of interest. For a few simple cases where analytical solutions exist, these solutions are found to be in complete agreement with our numerical results. We also analyze the distribution of fields and forces in more complex systems, and discuss the relevance of our findings to experimental observations. In particular, we demonstrate the single-beam trapping of a dielectric micro-sphere immersed in a liquid under conditions that are typical of optical tweezers.

## 1. Introduction

In a previous paper [1] we showed that the direct application of the Lorentz law of force in conjunction with Maxwell's equations can yield a complete picture of the electromagnetic force in metallic as well as dielectric media. In the case of the dielectrics, bound charges and bound currents were found to be responsible, respectively, for the electric and magnetic components of the Lorentz force. When a dielectric medium is homogeneous and isotropic, or can be divided into two or more such regions – each with its own uniform dielectric constant $\varepsilon$ – the bound charges appear only at the surface(s) and/or the interface(s) between adjacent dielectrics. The bound currents, however, induced by the local *E*-field in proportion to the time-rate of change of the polarization $\boldsymbol{P} = \varepsilon_o(\varepsilon - 1)\boldsymbol{E}$, are distributed throughout the medium. The *E*-field of the light exerts a force on the induced charge density, while the local *H*-field exerts a force on the induced current density. The net force can then be obtained by integrating these forces over the entire volume of the dielectric.

The above method, although involving no approximations, differs from the so-called "rigorous" methods commonly used in computing the force of radiation [2-4]. The primary difference is that we sidestep the use of Maxwell's stress tensor by reaching directly to bound electrons and their associated currents, treating them as the localized sources of electromagnetic force under the influence of the light's *E*- and *B*-fields. We also avoid the use of the two-component approach, where scattering and gradient forces are treated separately, either in the geometric-optical regime of ray-tracing [5], or in the electromagnetic regime where the light's momentum and intensity gradient are tracked separately [6]. Unlike some of the other approaches, our method computes the force density distribution and not just the total force.

In this paper we use Finite-Difference-Time-Domain (FDTD) computer simulations to obtain the electromagnetic field distribution in and around several dielectric media. The theoretical underpinnings of our computational method are described in Section 2. Section 3 is devoted to comparisons with exact solutions in the case of a dielectric slab illuminated at normal incidence, and also in the case of a semi-infinite medium under a *p*-polarized plane-wave at oblique incidence. In Section 4 we show that a one-dimensional Gaussian beam propagating in an isotropic, homogeneous medium, exerts either an expansive or a compressive lateral force on its host medium, depending on the beam's polarization state. The case of a top-hat-shaped beam entering a semi-infinite dielectric medium at oblique incidence is also covered in Section 4. In Section 5 we study the behavior of a cylindrical glass rod illuminated by a (one-dimensional) Gaussian beam, paying particular attention to the effects of polarization on the force-density distribution. In Section 6 we analyze the single-beam trapping of a small spherical bead, immersed in a liquid, by a sharply focused laser beam; both linear and circular polarization states of the beam are shown to result in strong trapping forces. A dielectric half-slab under a one-dimensional Gaussian beam centered on one edge of the slab is studied in Section 7. General remarks and conclusions are the subject of Section 8.

**2. Theoretical considerations**

To compute the force of the electromagnetic radiation on a given medium, we solve Maxwell's equations numerically to determine the distributions of the *E*- and *H*-fields (both inside and outside the medium). We then apply the Lorentz law

$$\boldsymbol{F} = \rho_b \boldsymbol{E} + \boldsymbol{J}_b \times \boldsymbol{B}, \tag{1}$$

where $\boldsymbol{F}$ is the force density, and $\rho_b$ and $\boldsymbol{J}_b$ are the bound charge and current densities, respectively [7]. The magnetic induction $\boldsymbol{B}$ is related to the *H*-field via $\boldsymbol{B} = \mu_o \boldsymbol{H}$, where $\mu_o = 4\pi \times 10^{-7}$ henrys/meter is the permeability of free space. In the absence of free charges $\nabla \cdot \boldsymbol{D} = 0$, where $\boldsymbol{D} = \varepsilon_o \boldsymbol{E} + \boldsymbol{P}$ is the displacement vector, $\varepsilon_o = 8.8542 \times 10^{-12}$ farads/meter is the free-space permittivity, and $\boldsymbol{P}$ is the local polarization density. In linear media, $\boldsymbol{D} = \varepsilon_o \varepsilon \boldsymbol{E}$, where $\varepsilon$ is the medium's relative permittivity; hence, $\boldsymbol{P} = \varepsilon_o (\varepsilon - 1) \boldsymbol{E}$.

When $\nabla \cdot \boldsymbol{D} = 0$, the bound-charge density $\rho_b = -\nabla \cdot \boldsymbol{P}$ may be expressed as $\rho_b = \varepsilon_o \nabla \cdot \boldsymbol{E}$. Inside a homogeneous, isotropic medium, $\boldsymbol{E}$ being proportional to $\boldsymbol{D}$ and $\nabla \cdot \boldsymbol{D} = 0$ imply that $\rho_b = 0$; no bound charges, therefore, exist inside such media. However, at the interface between two adjacent media, the component of $\boldsymbol{D}$ perpendicular to the interface, $\boldsymbol{D}_\perp$, must be continuous. The implication is that $\boldsymbol{E}_\perp$ is discontinuous and, therefore, bound charges exist at such interfaces; the interfacial bound charges will thus have areal density $\sigma_b = \varepsilon_o (E_{2\perp} - E_{1\perp})$. Under the influence of the local *E*-field, these charges give rise to a Lorentz force density $\boldsymbol{F} = \frac{1}{2} \, Real(\sigma_b \boldsymbol{E}^*)$, where $\boldsymbol{F}$ is the force per unit area of the interface. Since the tangential *E*-field, $\boldsymbol{E}_{\|}$, is continuous across the interface, there is no ambiguity as to the value of $\boldsymbol{E}_{\|}$ that should be used in computing the force. As for the perpendicular component, the average $\boldsymbol{E}_\perp$ across the boundary, $\frac{1}{2}(\boldsymbol{E}_{1\perp} + \boldsymbol{E}_{2\perp})$, must be used in calculating the interfacial force [1,8].

The only source of electrical currents within dielectric media are the oscillating dipoles, their (bound) current density $\underline{\boldsymbol{J}}_b$ in a dispersionless medium being given by

$$\underline{\boldsymbol{J}}_b = \partial \underline{\boldsymbol{P}}/\partial t = \varepsilon_0(\varepsilon - 1)\partial \underline{\boldsymbol{E}}/\partial t. \qquad (2)$$

(In this and the following equations, the underline denotes time dependence; symbols without the underline are used to represent the complex amplitudes of time-harmonic fields.) Assuming time-harmonic fields with the time-dependence factor $\exp(-i\omega t)$, one can rewrite Eq. (2) as $\boldsymbol{J}_b = -i\omega\varepsilon_0(\varepsilon - 1)\boldsymbol{E}$. The $B$-field of the electromagnetic wave exerts a force on the bound current according to the Lorentz law, namely, $\boldsymbol{F} = \frac{1}{2}Real\,(\boldsymbol{J}_b \times \boldsymbol{B}^*)$, where $\boldsymbol{F}$ is the (time-averaged) force per unit volume.

In FDTD simulations involving dispersive media, specific models (i.e., Debye, Drude, or Lorentz models) are used to represent the frequency dependence of the dielectric function $\varepsilon(\omega)$. Maxwell's equations are integrated in time over a discrete mesh, with the (model-dependent) dispersive behavior of the medium cast onto a time-dependent polarization vector. We assume $\varepsilon(\omega) = \varepsilon_\infty + \Delta\varepsilon(\omega)$, where the real-valued $\varepsilon_\infty$ denotes the relative permittivity of the medium in the absence of dispersion; the effects of dispersion enter through the complex-valued function $\Delta\varepsilon(\omega)$. The fraction of the polarization current density that embodies the contribution of $\Delta\varepsilon(\omega)$ is denoted by $\underline{\boldsymbol{J}}_p$. As the FDTD simulation progresses, determining the $E$- and $H$-fields as functions of time, $\underline{\boldsymbol{J}}_p$ is computed concurrently by solving the relevant differential equations of the chosen model.

To derive an expression for the total (i.e., free carrier + bound) current density $\underline{\hat{\boldsymbol{J}}}$ in a generally absorbing and dispersive medium, we begin with the following Maxwell equation:

$$\nabla \times \underline{\boldsymbol{H}} = \sigma \underline{\boldsymbol{E}} + \partial \underline{\boldsymbol{D}}/\partial t. \qquad (3)$$

Here the real-valued $\sigma$ denotes the material's conductivity for the free carriers. In our simulations, $\sigma$ is assumed to be a constant, independent of the frequency $\omega$; that is, the free carriers' conductivity is assumed to be dispersionless. Note that setting $\sigma = 0$ does not necessarily guarantee a transparent medium, as absorption could enter through the imaginary component of the complex permittivity, $\Delta\varepsilon(\omega)$.

Substituting for $\underline{\boldsymbol{D}}$ in Eq. (3) from the constitutive relation $\underline{\boldsymbol{D}} = \varepsilon_0 \underline{\boldsymbol{E}} + \underline{\boldsymbol{P}}$, we arrive at

$$\varepsilon_0\,\partial \underline{\boldsymbol{E}}/\partial t = \nabla \times \underline{\boldsymbol{H}} - (\sigma \underline{\boldsymbol{E}} + \partial \underline{\boldsymbol{P}}/\partial t) = \nabla \times \underline{\boldsymbol{H}} - \underline{\hat{\boldsymbol{J}}}, \qquad (4)$$

where, by definition, the total current density $\underline{\hat{\boldsymbol{J}}}$ is given by

$$\underline{\hat{\boldsymbol{J}}} = \nabla \times \underline{\boldsymbol{H}} - \varepsilon_0\,\partial \underline{\boldsymbol{E}}/\partial t. \qquad (5)$$

In FDTD computations, Eq. (3) is often rearranged as follows:

$$\varepsilon_0\varepsilon_\infty\,\partial \underline{\boldsymbol{E}}/\partial t = \nabla \times \underline{\boldsymbol{H}} - \sigma \underline{\boldsymbol{E}} - \underline{\boldsymbol{J}}_p, \qquad (6)$$

where, for dispersive media (e.g., Debye, Drude, and Lorentz models), $\varepsilon_\infty$ represents the frequency-independent part of the relative permittivity, while $\underline{\boldsymbol{J}}_p$ models dispersion ($\underline{\boldsymbol{J}}_p = 0$ for non-dispersive dielectrics). When computing $\underline{\hat{\boldsymbol{J}}}$, it is convenient to eliminate the time-derivative of $\underline{\boldsymbol{E}}$ from Eq. (5), as only one time-slice of $\underline{\boldsymbol{E}}$ is normally stored during FDTD computations. From Eq. (6), $\varepsilon_0\partial \underline{\boldsymbol{E}}/\partial t = (\nabla \times \underline{\boldsymbol{H}} - \sigma \underline{\boldsymbol{E}} - \underline{\boldsymbol{J}}_p)/\varepsilon_\infty$, which, when substituted in Eq. (5), yields

$$\underline{\hat{\boldsymbol{J}}} = (\sigma \underline{\boldsymbol{E}} + \underline{\boldsymbol{J}}_p)/\varepsilon_\infty + (1 - 1/\varepsilon_\infty)\nabla \times \underline{\boldsymbol{H}}. \qquad (7)$$

Computing the total current density $\underline{\hat{\boldsymbol{J}}}$ (i.e., the sum of the conduction and polarization current densities) at a given instant of time thus requires only the contemporary values of $\underline{\boldsymbol{E}}$, $\underline{\boldsymbol{H}}$, and $\underline{\boldsymbol{J}}_p$, which are readily available during the normal progression of an FDTD simulation.

Given the electromagnetic fields $\underline{E}$ and $\underline{H}$ as functions of spatial coordinates and time, we compute the force density distribution by time-averaging the Lorentz equation, namely,

$$<\underline{F}> = (1/T)\int_0^T (\underline{E}\,\nabla\cdot\varepsilon_o\underline{E} + \underline{\hat{J}}\times\mu_o\underline{H})\,dt. \qquad (8)$$

The time integral in the above equation is taken over one period of the (time-harmonic) light field. Contributions to the Lorentz force by the electric-field, $\rho\underline{E}$ (where the total charge density $\rho = \nabla\cdot\varepsilon_o\underline{E}$), and by the magnetic-field, $\underline{J}\times\underline{B}$, can be readily identified in Eq. (8).

## 3. Comparison to exact solutions

The case of a dielectric slab of index $n$ and thickness $d$ illuminated by a plane-wave at normal incidence was analyzed in our previous paper [1]. To fix the frame of reference, we assume that the slab is parallel to the $xy$-plane and the beam propagates in the $-z$ direction. At normal incidence, there are no induced charges, and the electromagnetic pressure in its entirety may be attributed to the magnetic component of the Lorentz force. Inside the slab, a pair of counter-propagating plane-waves (along $\pm z$) interfere with each other and set up a system of fringes. The force is distributed within the fringe pattern, being positive (i.e., in the $+z$ direction) over one-half of each fringe and negative over the other half. The net force is obtained by integrating the local force density through the thickness of the slab.

Figure 1 shows the computed force density $F_z$ inside a slab illuminated at normal incidence by a plane-wave of wavelength $\lambda_o = 640$nm and $E$-field amplitude $E_o = 1.0$ V/m. (Computed values of $F_x$ and $F_y$ were zero, as expected.) The slab is suspended in free-space, and has refractive index $n = 2.0$ and thickness $d = 110$nm. The total force along the $z$-axis (per unit cross-sectional area) is found to be $\int F_z(z)\,dz = -2.481$ pN/m$^2$ ($\Delta z = 5.0$ nm in our FDTD simulations) versus the exact value of $-2.479$ pN/m$^2$, obtained from theoretical considerations [1]. For a quarter-wave-thick slab, $d = 80$nm, the simulation yields $\int F_z(z)\,dz = -3.192$ pN/m$^2$, where the exact solution is $-3.188$ pN/m$^2$.

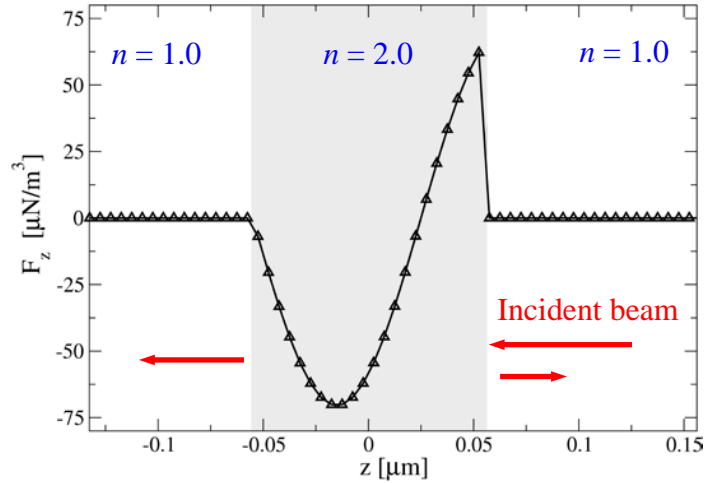

Fig. 1. Computed force density $F_z$ (per unit cross-sectional area) versus $z$ inside a dielectric slab illuminated with a normally incident plane-wave ($\lambda_o = 0.64$ μm). The slab, suspended in free-space, has $n = 2.0$, $d = 110$nm. The incident beam propagates along the negative $z$-axis.

As another example, consider a $p$-polarized plane-wave ($\lambda_o = 650$ nm) incident at $\theta = 50°$ on a semi-infinite dielectric of refractive index $n = 3.4$, located in the region $z < 0$. Figure 2 shows computed time-snapshots of the $E_z$ component of the field. In Fig. 2(a) the transition of the refractive index from $n_o = 1.0$ to $n_1 = 3.4$ is abrupt, and the force per unit area due to the induced surface charges, computed in FDTD with $\Delta y = \Delta z = 5.0$ nm, is $(F_y, F_z) = (1.696, 2.468)$ pN/m$^2$, versus the exact value of $(1.699, 2.467)$ pN/m$^2$. For $n = 2.0$, FDTD simulations yield $(F_y, F_z) = (1.5903, 1.6475)$ pN/m$^2$ versus the exact value of $(1.5911, 1.6493)$ pN/m$^2$. Figure 2(b) corresponds to a semi-infinite dielectric with a rapid (linear) transition of the refractive index from $n_o = 1.0$ to $n_1 = 3.4$ over a 40 nm-thick region. In this case, the force per unit surface area computed in FDTD is $(F_y, F_z) = (1.744, 2.638)$ pN/m$^2$.

We mention in passing that the case of a finite-diameter beam at Brewster's incidence on a dielectric wedge is also amenable to analytical as well as numerical solution, and that the numerically computed forces are in excellent agreement with the theoretical values [8].

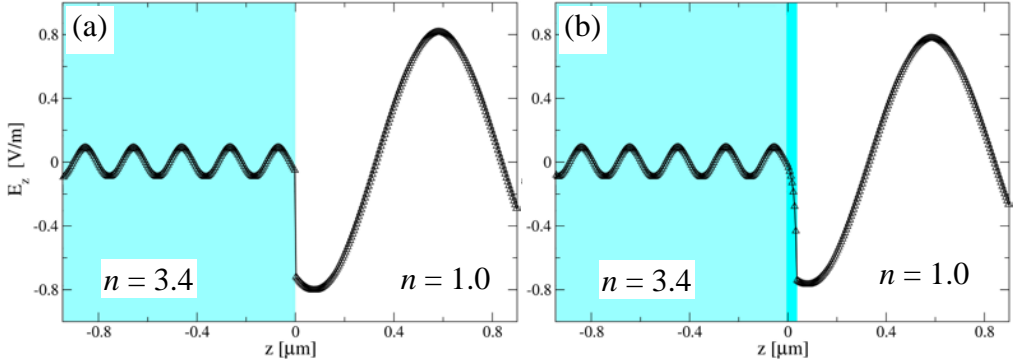

Fig. 2. Time-snapshots of the $E_z$ component of a $p$-polarized plane-wave, $\lambda_o = 0.65$ μm, incident at $\theta = 50°$ on a semi-infinite dielectric of refractive index $n = 3.4$, located in the region $z < 0$. (a) Abrupt transition of the refractive index at $z = 0$. (b) Linear transition of the refractive index from $n_o = 1.0$ to $n_1 = 3.4$ over a 40 nm-thick region.

The above results establish the accuracy of our numerical calculations, especially when the surface charge is limited to a single pixel, as was the case in Fig. 2(a). Making smooth transitions at the boundaries between regions of differing refractive indices, as was done in the case of Fig. 2(b), is a tool that is available in numerical simulations. Such smooth transitions at the boundaries may represent physical reality, or they may be used as an artificial tool to eliminate sharp discontinuities and singularities of the equations. To the extent that such smoothing operations do not modify the actual physics of the problem under consideration, they may be used with varying degrees of effectiveness.

## 4. Force exerted by beam's edge on the host medium

We have shown in [1] that, among other things, the magnetic Lorentz force is responsible for a lateral pressure exerted on the host medium at the edges of a finite-diameter beam; the force per unit area at each edge (i.e., side-wall) of the beam is given by

$$F^{(\text{edge})} = \tfrac{1}{4}\varepsilon_o(\varepsilon - 1)|E_o|^2. \tag{9}$$

Here $|E_o|$ is the magnitude of the $E$-field of a (finite-diameter) plane-wave in a medium of dielectric constant $\varepsilon$. If the $E$-field is parallel (perpendicular) to the beam's edge, the force is compressive (expansive); in other words, the opposite side-walls of the beam tend to push the medium toward (away from) the beam center. The "edge force" does not appear to be sensitive to the detailed structure of the beam's edge; in particular, a one-dimensional Gaussian beam exhibits the edge force described by Eq. (9) when its (magnetic) Lorentz force on the host medium is integrated laterally on either side of the beam's center [1].

Consider a one-dimensional Gaussian beam (uniform along $x$, Gaussian along $y$, and propagating in the negative $z$-direction) in a homogeneous host medium of refractive index $n = 2.0$; the free-space wavelength of the cw beam is $\lambda_o = 0.65$ μm. Figure 3 shows time snapshots of the field profile (first row) and time-averaged force density distributions (second row) in the $yz$ cross-sectional plane. The beam's full-width at half-maximum-amplitude measured at the waist (located at $z = +0.3$ μm) is FWHM = 1.5 μm.

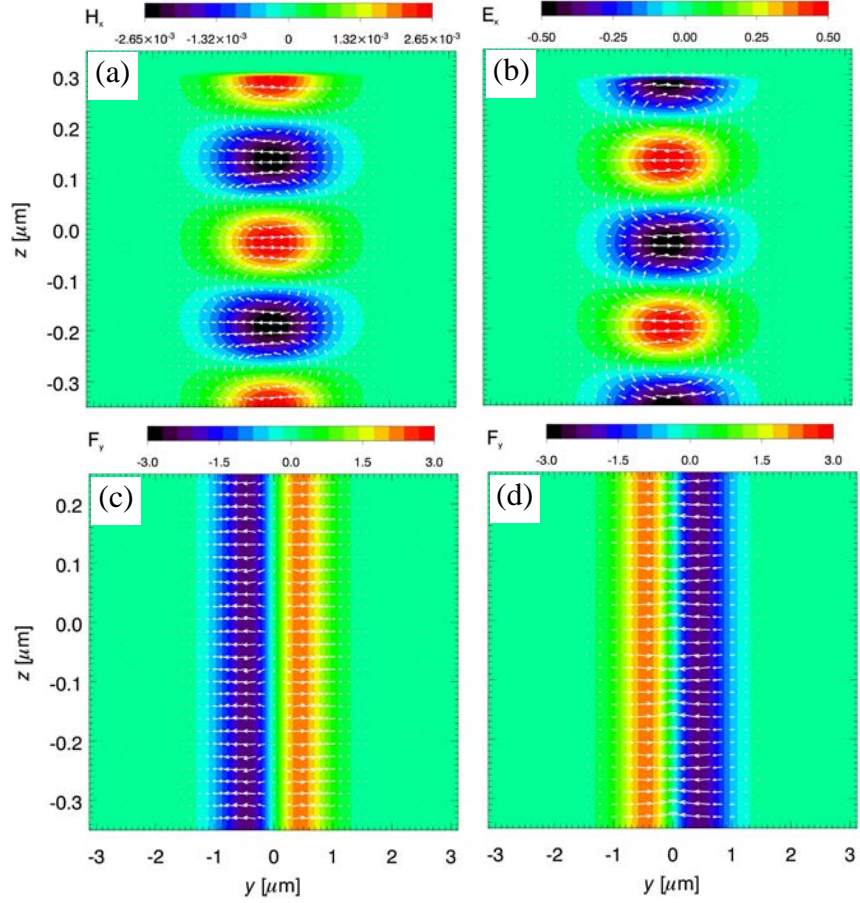

Fig. 3. One-dimensional Gaussian beam in a homogeneous medium of refractive index $n = 2.0$. The beam waist is at $z = 0.3$ μm, and the beam's propagation direction is along the negative $z$-axis. (a) Time snapshot of $H_x$ distribution for $p$-light, with superposed arrows depicting the $(E_y, E_z)$ vector field. (b) Time snapshot of $E_x$ distribution for $s$-light, with the $(H_y, H_z)$ vector field superposed. (c) Force density distribution of the $F_y$ component in the case of $p$-polarization, with the $(F_y, F_z)$ vector field superposed. (d) Distribution of the force density component $F_y$ for the $s$-polarized beam, with the $(F_y, F_z)$ vector field superposed.

In Fig. 3 the field is $p$-polarized on the left- and $s$-polarized on the right-hand-side. The upper-left frame shows a color-coded plot of the $H_x$ distribution, on which the $E$-field vector $(E_y, E_z)$ is superposed. The frame below it shows the force density plot of $F_y$, together with the $(F_y, F_z)$ vector field in the case of $p$-polarization. The upper-right frame is a time snapshot of the $E_x$ distribution, with the $H$-field vector $(H_y, H_z)$ superposed. The frame below it shows $F_y$ for the $s$-polarized beam, with the $(F_y, F_z)$ vector field superposed. These force-density fields are computed using time-averages (over one period of the optical wave, $T = \lambda_o/c$) of the time- and space-dependent force density. Note that $F_y$ is expansive for $p$-light and compressive for $s$-light. The integrated lateral force per unit area on each half of the beam, $\int F(x, y, z)\,\mathrm{d}y$, is 1.6577 pN/m$^2$ for $p$-light and 1.6581 pN/m$^2$ for $s$-light. The theoretical value of the force per unit area on each edge of the beam given by Eq. (9) with $E_o = 0.5$ V/m is $F^{(edge)} = \pm 1.66$ pN/m$^2$.

Another example of the same phenomenon (lateral force at the beam's edge) is exhibited by a finite-diameter beam that enters a semi-infinite dielectric at oblique incidence. Figure 4 shows time-snapshots of the field distributions for a linearly-polarized wave ($\lambda_o = 0.65\mu m$) having a top-hat cross-sectional profile with smooth edges. The beam is incident from the free-space at $\theta_{inc} = 50°$ onto a semi-infinite dielectric of index $n_s = 2.0$; the dielectric fills the half-space $z < 0$. The case of $p$-polarization is shown on the left-, that of $s$-polarization on the right-hand side. For the $p$-polarized beam, its angle of incidence being close to Brewster's angle $\theta_B = 63.43°$, there is little reflectivity at the interface and most of the light is transmitted through to the dielectric medium, whereas for the $s$-polarized beam a good fraction of the incident light is being reflected.

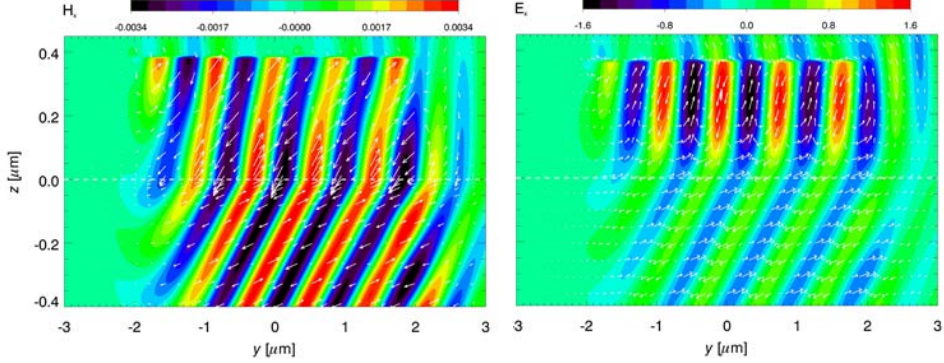

Fig. 4. Time snapshots of the field components for a linearly-polarized wave ($\lambda_o = 0.65\mu m$) having a top-hat cross-sectional profile with smooth edges, incident at $\theta_{inc} = 50°$ from free-space onto a semi-infinite dielectric of refractive index $n_s = 2.0$, located in the half-space $z < 0$. (Left) Magnetic field $H_x$ in the case of $p$-polarization; the superposed arrows represent the electric-field ($E_y$, $E_z$). (Right) Electric field $E_x$ in the case of $s$-polarization; the superposed arrows represent the magnetic-field ($H_y$, $H_z$).

Computed force densities ($F_y$, $F_z$) inside the dielectric medium are displayed in Fig. 5; the interface region has been excluded to avoid, in the case of $p$-light, the high force region of the induced surface charges. (For $s$-light the $E$-field is continuous at the boundary and, therefore, no surface charges are induced.) In both cases the force fields near the leading edge of the beam show oscillatory behavior, whereas the trailing edge is fairly smooth. Despite oscillations near the edge, the force is seen to be generally expansive for $p$-light and compressive for $s$-light, that is, the sign of the integrated lateral force on each side of the center is consistent with the theoretical arguments presented in [1]. In units of pN/m$^2$, the integrated force ($\int F_y \, dy$, $\int F_z \, dy$) for $p$-light is (−2.64, −1.13) for the left-edge and (2.59, 1.096) for the right-edge. These edge forces are nearly identical in strength (to better than ±1.5%), are orthogonal to the propagation direction within the dielectric, and are in fair agreement with the theoretical value of ±(2.51, 1.04) pN/m2 obtained from Eq. (9) with $E_o = 0.64$ V/m. The corresponding edge forces for the $s$-light depicted in Fig. 5, right-hand column, are (1.957, 0.821) for the left edge and (−1.939, −0.815) for the right edge. Again, this compressive force is orthogonal to the propagation direction, and is in reasonable agreement with the theoretical value of ±(1.92, 0.8) pN/m2 obtained from Eq. (9) with $E_o = 0.56$ V/m.

## 5. Cylindrical rod illuminated by Gaussian beam

Figure 6 shows time-snapshot plots of $H_x$, $E_y$, $E_z$ components of a $p$-polarized, one-dimensional Gaussian beam ($\lambda_o = 0.65$ μm, amplitude FWHM = 0.5 μm), propagating in free-space along the negative $z$-direction; the beam's waist is at $z = 0.5$ μm. Given the amplitude profile as $H_x(y, z = 0.5\mu m) = H_o \exp[-(y/y_o)^2]$, the value of $y_o$ at the waist is 0.3μm, which corresponds to a divergence half-angle $\theta = \arctan(\lambda_o/\pi y_o) \approx 35°$. The small diameter of the beam at the waist thus results in its rapid divergence along the propagation direction. The cone angle of the beam, although large enough to exhibit lateral trapping of small dielectric objects, is not sufficient to produce vertical trapping [5], as will be seen below.

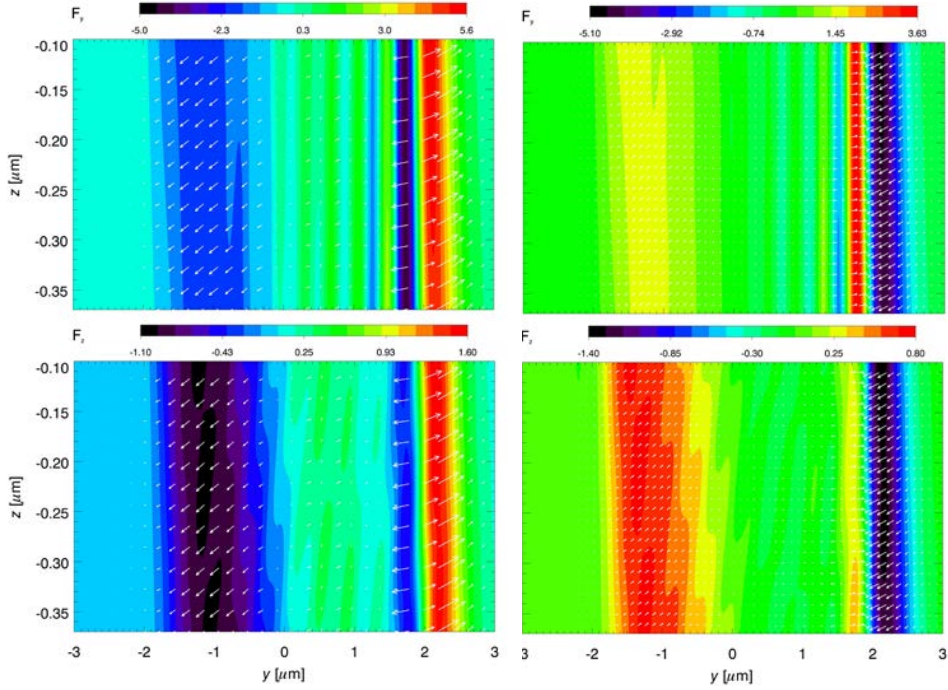

Fig. 5. Left column: $F_y$ and $F_z$ force density plots for the *p*-polarized beam of Fig. 4(a). Right column: $F_y$ and $F_z$ force fields for the *s*-polarized beam of Fig. 4(b). The free-space/dielectric interface is not shown to exclude from the color scale the region of high force density due to the induced surface charges in the case of *p*-polarization.

In our simulations the initial *H*-field amplitude was $H_o = 2.6544$ A/m, leading to an initial *E*-field amplitude (at the center of the Gaussian) $E_o = Z_o H_o = 1000$ V/m; here $Z_o = \sqrt{\mu_o/\varepsilon_o}$ is the impedance of free-space. Ignoring for the moment the complication that the *E*-field consists of both $E_y$ and $E_z$, the *z*-component of the beam's Poynting vector is approximated as

$$S_z(y, z = 0.5\mu\text{m}) = \tfrac{1}{2} E_o H_o \exp[-2(y/y_o)^2]. \tag{10}$$

The incident beam's integrated power (per unit length along the *x*-axis) is then found to be

$$\int S_z(y, z = 0.5\mu\text{m}) \mathrm{d}x = \sqrt{\pi/8}\, E_o H_o\, y_o = 0.5 \times 10^{-3}\ \text{W/m}. \tag{11}$$

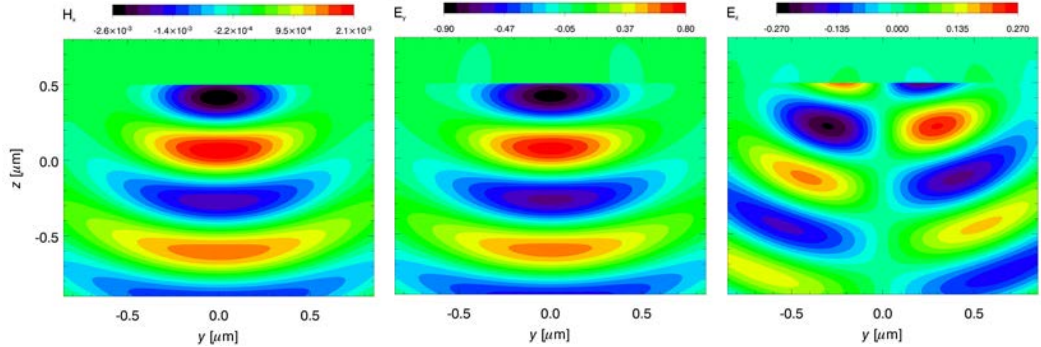

Fig. 6. Left to right: time-snapshot plots of $H_x$, $E_y$, $E_z$ components of a *p*-polarized, one-dimensional Gaussian beam ($\lambda_o = 0.65$ μm, FWHM = 0.5 μm), propagating in free-space in the negative *z*-direction; the beam's waist is at $z = 0.5$ μm.

In this Gaussian beam's path we placed a cylindrical rod (radius $r_c = 0.2$ μm, index $n_c = 2.0$) parallel to the *x*-axis at $(y, z) = (0.2$ μm, $0)$. The cylinder's index changes linearly at the surface, from $n = 1.0$ to $2.0$, within a 20 nm-thick transition layer (grid cell size = 5nm). Figure 7, left column, shows computed plots of $|H_x|$ as well as force-density components $F_y$ and $F_z$ for this *p*-polarized beam. The integrated force is $(\int F_y(y, z)\,dydz, \int F_z(y, z)\,dydz) = (0.17, 0.21)$ pN/m without including the surface-charge contribution, and $(-0.4, -0.58)$ pN/m when the surface-charge contribution is included. The net force is thus pulling the cylinder to the left while, at the same time, pushing it downward.

The right-hand column of Fig. 7 shows the case of a *s*-polarized Gaussian beam (same parameters as in Fig. 6) illuminating the same glass cylinder as above. From top to bottom, the figure shows computed plots of $|E_x|$ and force-density components $F_y$ and $F_z$. Here surface charges are absent, and the force is purely due to the magnetic component of the Lorentz force. The integrated force is $(\int F_y(y, z)\,dydz, \int F_z(y, z)\,dydz) = (-0.14, -0.68)$ pN/m, which, as in the case of *p*-light, tends to pull the rod to the left while pushing it downward. The pull to the left is consistent with what has been observed in optical trapping experiments, but the downward push is contrary to the known behavior of single-beam traps (i.e., optical tweezers) [9,10]. We will see in the next section that vertical trapping benefits from the submersion of the particle in a liquid, where the index mismatch between the dielectric and its environment is reduced and the scattering force (caused by surface reflections) is minimized. More importantly, the results of the next section demonstrate the necessity of a sharper focus (i.e., a smaller beam waist diameter) [5] than has been used in the present simulations.

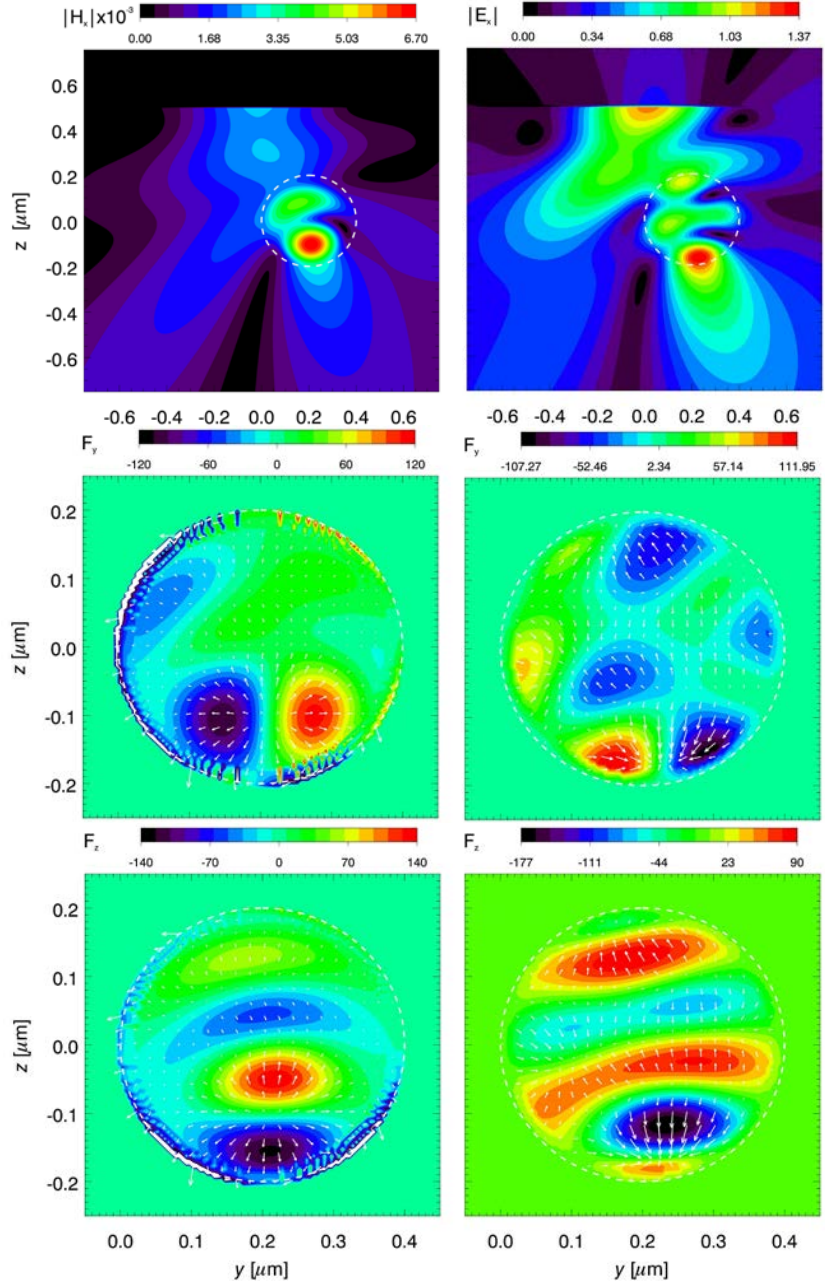

Fig. 7. Gaussian beam of Fig. 6 in the presence of a glass cylinder having $r_c = 0.2$ μm, $n_c = 2.0$, centered at $(y, z) = (0.2$ μm$, 0)$. Left column, top to bottom: distributions of $|H_x|$, $F_y$, $F_z$ for $p$-light. Force density profiles include the contributions by surface charges. Right column, top to bottom: distributions of $|E_x|$, $F_y$, $F_z$ for $s$-light. The field plots at the top are shown over a large region; dashed lines indicate the boundary of the cylinder.

## 6. Spherical bead immersed in liquid and trapped by a focused beam

In this section we present the distribution of force in and around a spherical particle of radius $r_s = 1.0\mu m$ and index $n_s = 1.6$ immersed in a liquid of index $n_o = 1.3$. The particle is displaced along the $x$-axis, its center being at $(x, y, z) = (\Delta x, 0, 0)$. The incident beam, obtained by focusing a $\lambda_o = 0.65\mu m$ plane-wave through a 1.25NA immersion objective, is sourced at $(x, y, z) = (0, 0, 1.2\mu m)$ and propagates in the negative $z$-direction. The beam entering the objective's pupil may be circularly or linearly polarized. The actual point of focus can be placed at various locations along the $z$-axis, $(x, y, z) = (0, 0, \Delta z)$, inside or outside the spherical particle. The lens is designed for diffraction-limited focusing within the immersion liquid and, in the absence of the spherical particle, the focused spot is free from aberrations and has a total power (i.e., integrated $S_z$ within the focal plane) of $P = 0.5$ mW.

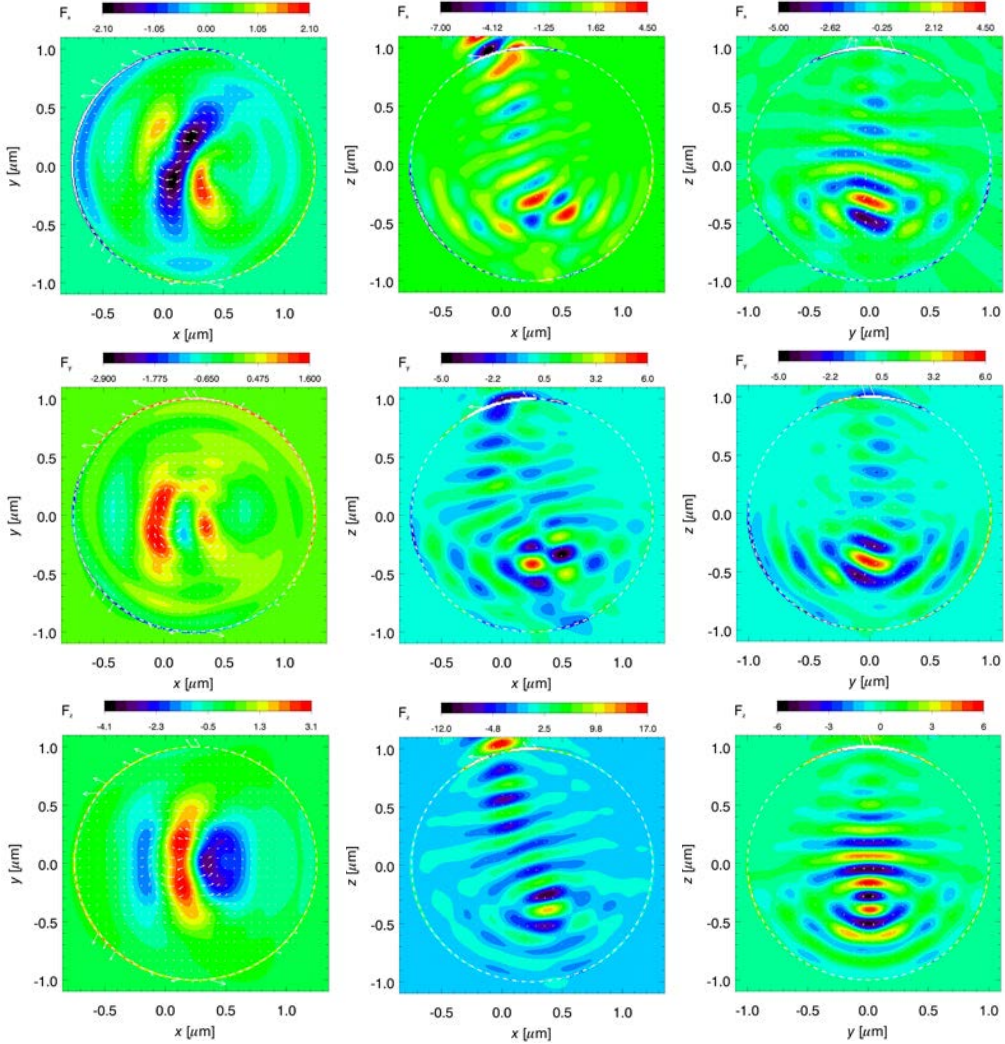

Fig. 8. Cross-sectional plots of force-density component distributions (color contours) through the center of dielectric sphere ($r_s = 1.0$ μm, $n_s = 1.6$), centered at $(x, y, z) = (0.25\mu m, 0, 0)$ and immersed in a medium of index $n_o = 1.3$. The arrows show the projection of the force density vector in the cross-sectional plane, e.g., plots on the $yz$-plane show the $(F_y, F_z)$ vector field. Top to bottom: $F_x, F_y, F_z$. Left to right: $xy$-plane, $xz$-plane, $yz$-plane. The circularly-polarized incident beam is focused at $(x, y, z) = (0, 0, 0.7\mu m)$ through a 1.25NA immersion objective.

Computing the force of radiation on an immersed solid particle requires that the bound charges on the surface of the particle be distinguished from the charges induced on the surrounding liquid. This issue has been discussed in detail in [8], where we have also argued that the perpendicular component of the *E*-field at the particle's surface – used in computing the *E*-field contribution to the Lorentz force – may be obtained in several different ways. In the present simulations we used the *average E*-field across the boundary (i.e., at the sphere surface) in computing the Lorentz force on the interfacial charges accumulated on the sphere.

For the case of focusing a circularly-polarized beam at $(x, y, z) = (0, 0, 0.7\mu m)$ onto a particle centered at $(0.25\mu m, 0, 0)$, Fig. 8 shows color-coded force-component distributions in various cross-sectional planes through the center of the sphere. The superposed arrows show the projection of the force-density vector in the corresponding plane, e.g., plots on the *yz*-plane show the $(F_y, F_z)$ vector field. Contributions to force-density come from both (bound) surface charges and (bound) electric currents within the volume of the sphere. When the force density exceeds the range of color coding, the corresponding region is depicted as white. This happens in the area where the beam enters the sphere, seen in the *xz* and *yz* cross-sectional planes of Fig. 8 near the top of each frame. Reflection, refraction, and diffraction create complex interference patterns within the spherical particle, as evidenced by the richly textured force density profiles throughout the volume of the sphere. The surrounding liquid also experiences the effects of radiation pressure, as the bound currents within the liquid body feel the Lorentz force of the *B*-field, while the bound charges on the liquid side of the interface sense the Lorentz force of the *E*-field, in the same way that the particle itself experiences such forces. The force imparted to the liquid is strongest in the vicinity of the region where the beam enters the particle, but weak forces are felt (by the liquid) also in those regions where the light leaves the solid particle and re-emerges into the liquid environment.

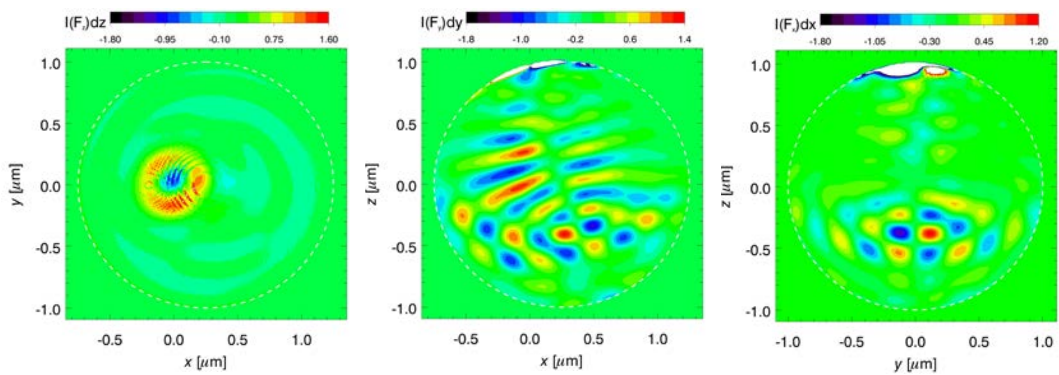

Fig. 9. Cross-sectional profiles of the integrated force-density components along the direction perpendicular to each cross-section. (Left) Distribution of $\int F_z(x, y, z)\,dz$ in the central *xy*-plane. (Middle) Distribution of $\int F_y(x, y, z)\,dy$ in the central *xz*-plane. (Right) Distribution of $\int F_x(x, y, z)\,dx$ in the central *yz*-plane. The range of integration includes the charges on the surface of the sphere, but excludes any forces that might be exerted on the surrounding liquid.

For the system of Fig. 8, profiles of the integrated force density in the normal direction to each cross-section are shown in Fig. 9; the distribution of $\int F_z(x, y, z)\,dz$ in the central *xy*-plane appears on the left, that of $\int F_y(x, y, z)\,dy$ in the central *xz*-plane in the middle, and the profile of $\int F_x(x, y, z)\,dx$ in the central *yz*-plane on the right-hand side. The range of integration includes the (bound) charges on the surface of the particle, but excludes any forces that might be exerted by the entering/exiting light on the surrounding liquid. The presence of a strong force on the surface charges is indicated by the white regions (i.e., exceeding the color scale) near the top of the frame in the *xz* and *yz* cross-sections depicted in Fig. 9.

In the above case of focusing a circularly-polarized beam at $(x, y, z) = (0, 0, 0.7\mu m)$ onto a glass bead centered at $(0.25\mu m, 0, 0)$, the net force on the bead, obtained by integrating the force-density through the spherical volume, is $\iiint \boldsymbol{F}(x,y,z)\,\mathrm{d}x\,\mathrm{d}y\,\mathrm{d}z = (-0.40, +0.018, +0.34)$ pN. The direction of the total force clearly indicates trapping of the particle by the focused beam, as $F_x$ tends to pull the particle to the left while $F_z$ tries to lift it up, toward the center of the focused spot. (Similar calculations were carried out for different positions of the bead relative to the focused spot, and trapping was demonstrated in all cases; see Fig. 10.) It must be pointed out that the $y$-component of the net force, although small, is not negligible, thus indicating the transfer of a certain amount of angular momentum from the circularly polarized beam to the illuminated particle (this point will be discussed in more detail below).

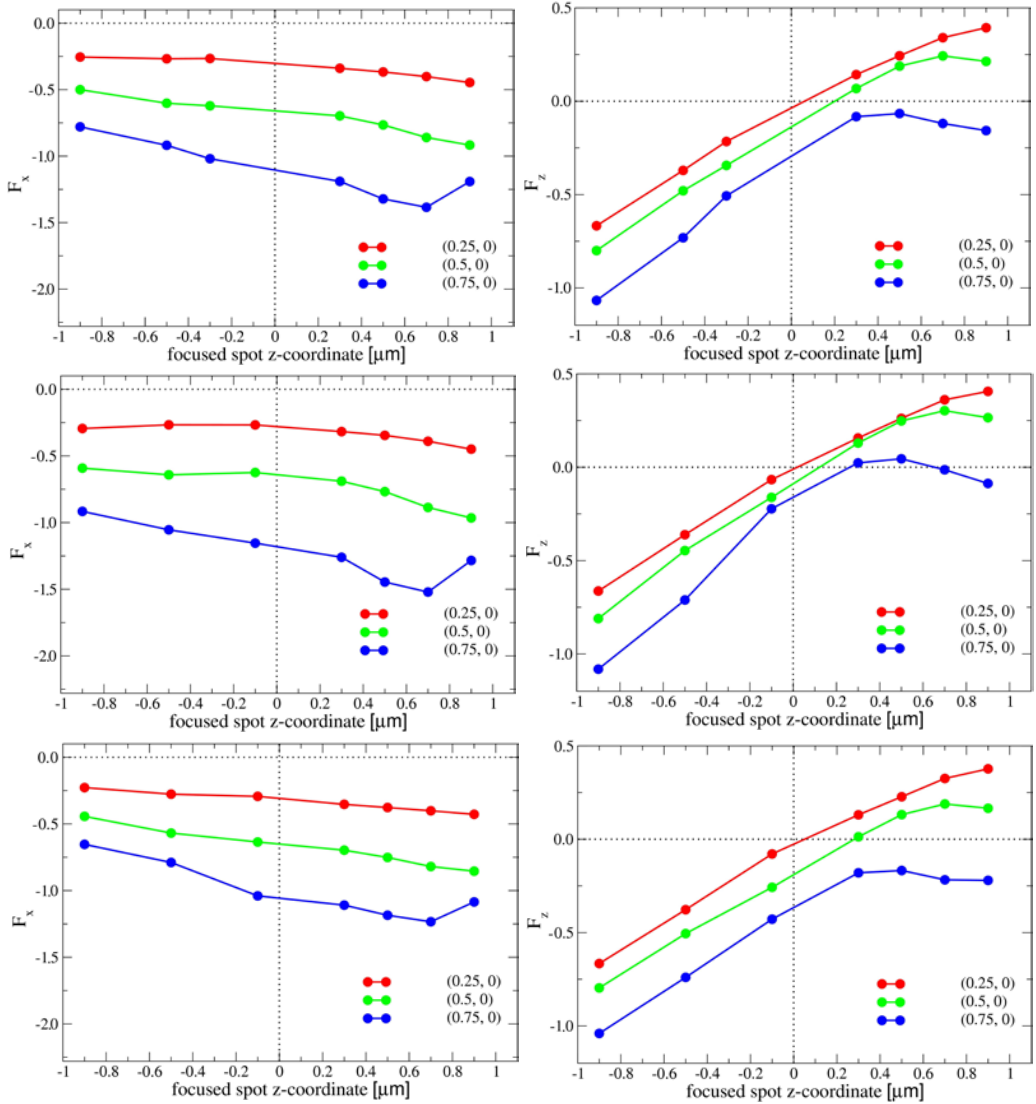

Fig. 10. Computed force components $F_x$ (left column) and $F_z$ (right column) versus the vertical displacement $\Delta z$ of the focal point from the sphere center ($\Delta z > 0$ when the beam is focused above the sphere center). $F_x$ and $F_z$ are in units of pico Newtons. Each colored curve represents a fixed lateral displacement $\Delta x$ of the center of the particle from the optical axis of the objective lens. The beam is focused in the liquid through a 1.25NA immersion lens. The polarization state of the 0.5 mW plane wave ($\lambda_o = 0.65\mu m$) entering the objective's pupil is: circular (top row), linear along the $x$-axis (middle row), linear along the $y$-axis (bottom row).

Plots of the total force components ($F_x$, $F_z$) versus vertical displacement $\Delta z$ of the focus center from the sphere center are shown in Fig. 10 ($\Delta z > 0$ when the focused spot is above the sphere center). Different colored curves correspond to different lateral displacements $\Delta x$ of the particle from the optical axis of the lens (red: $\Delta x = 0.25\mu m$, green: $\Delta x = 0.5\mu m$, blue: $\Delta x = 0.75\mu m$). The top row in Fig. 10 corresponds to a circularly polarized plane-wave entering the pupil of the lens. The middle row corresponds to linear polarization along $x$, and the bottom row represents the case of linear polarization along the $y$-axis. Note that for $\Delta z > 0$, $F_z$ is positive only when $\Delta x$ is small; this indicates that the particle is first trapped laterally by the (fairly strong) $F_x$, then lifted upward until its center nearly coincides with the center of the focused beam.

For linearly polarized light (whether along the direction of the lateral displacement of the bead, $x$, or perpendicular to that direction, $y$), $F_y$ was found to be zero, as expected from symmetry considerations. The non-zero values of $F_y$ obtained for circularly polarized light indicate the transfer of a certain amount of angular momentum from the beam to the particle. We found $F_y$ to be generally an order of magnitude weaker than $F_x$, fairly independent of $\Delta z$, and an increasing function of $\Delta x$. When the sense of polarization was reversed (say, from right- to left-circular), $F_y$ switched sign, while its magnitude – as well as the signs and magnitudes of $F_x$ and $F_z$ – remained unchanged. We thus believe the weak but non-zero values of $F_y$, predicted to exist under circularly-polarized light, are real and should be subjected to experimental verification.

We have also noticed that the inclusion of a thin layer of liquid on the surface of the particle results in somewhat stronger trapping forces. In practice, it is not unreasonable to expect a layer of liquid, with a thickness of several ten nanometers, to stick to the particle's exterior surface. The forces of radiation experienced by this liquid layer may thus have to be included in the overall force calculation in order to obtain more accurate results.

## 7. Dielectric half-slab illuminated near its side-wall

The case of a dielectric half-slab illuminated at one edge (i.e., side-wall) by a finite-diameter beam was briefly discussed in our previous paper [1] in conjunction with optical trapping experiments. We argued qualitatively that, even though the edge of a *p*-polarized beam inside the half-slab tends to exert an expansive force on its host medium (and hence drive it away from the beam's center), the net force experienced by the half-slab should be dominated by the surface charges induced on the slab's side-wall, which force tends to draw the slab toward the center of the beam. Attraction of the half-slab to the beam's center is thus expected to occur irrespective of whether the incident beam is *p*- or *s*-polarized.

In the present section we discuss this problem in some detail, using the one-dimensional Gaussian beam of Fig. 3 ($\lambda_o = 0.65\mu m$, FWHM = 1.5$\mu m$), normally incident from free-space onto a dielectric half-slab having $n_s = 2.0$ and thickness $d = 110$nm. The beam center and the slab-edge are assumed to coincide at $y = 0$. Figure 11 shows the computed field as well as force-density plots when the beam is *p*-polarized (left column) and *s*-polarized (right column). In the *p*-polarized case the color scale has been adjusted to exclude high force-density values at the edges/corners of the half-slab. The horizontal force component $F_y$ (second row) is mostly positive inside the slab but has very strong negative values on the slab's vertical edge, where there is a substantial charge accumulation. In the *s*-polarized case, no surface charges are induced (because the only *E*-field component, $E_x$, is everywhere parallel to the glass surfaces). The only forces in the case of *s*-polarization, therefore, are due to the bound currents within the medium, acted upon by the *H*-field of the light beam. Clearly, the force distributions are quite complex, and it is not immediately obvious if the total force in the horizontal direction is pulling the half-slab to the left or pushing it to the right; net (integrated) values of the force are thus needed to settle the question.

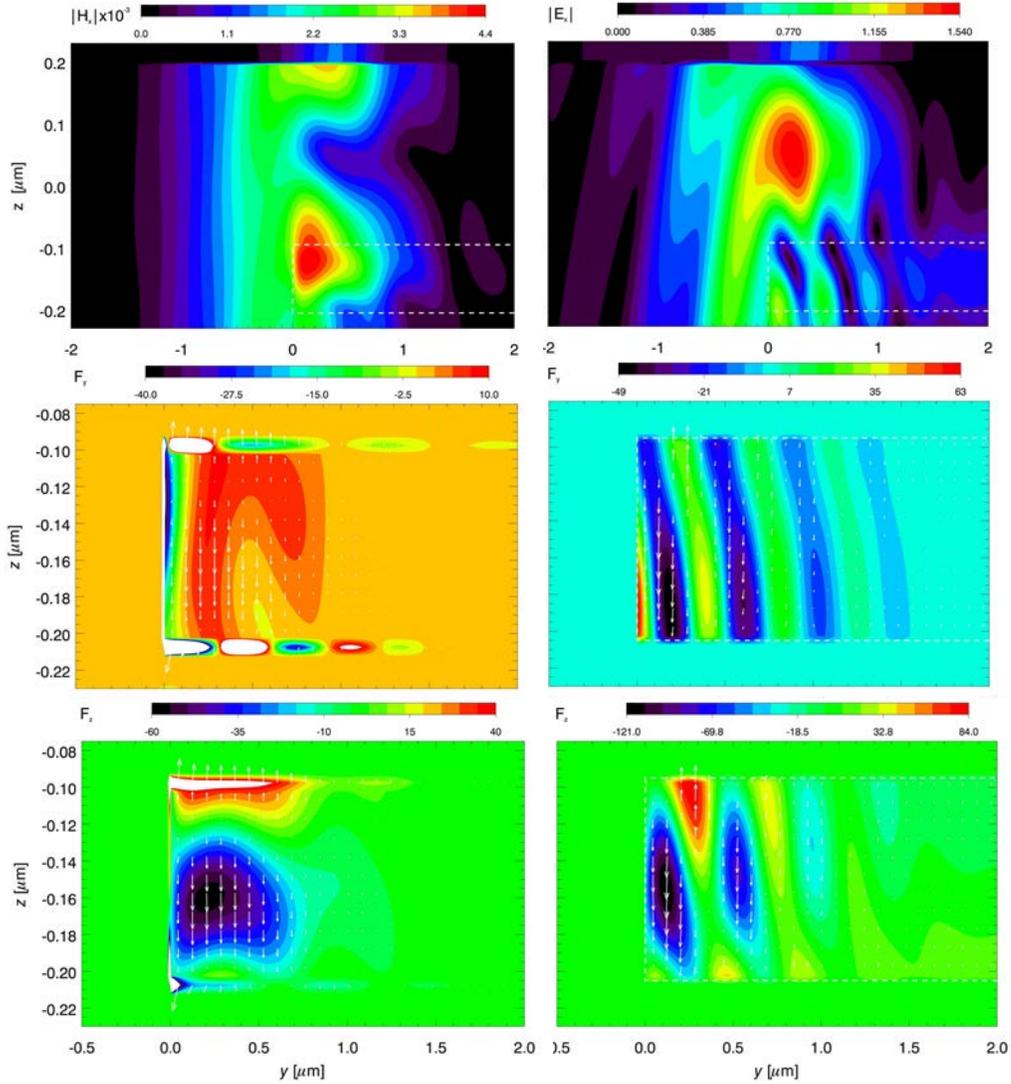

Fig. 11. Plots of field ($E_x$, $H_x$) and force-density ($F_y$, $F_z$) distribution when the Gaussian beam of Fig. 3 is incident from the free-space onto the edge (i.e., side-wall) of a dielectric half-slab. Left column: *p*-polarization; right column: *s*-polarization. Both the beam center and the slab-edge are at $y = 0$. The top row shows profiles of $H_x$ (*p*-light) and $E_x$ (*s*-light) over a large area that includes the slab's edge (dashed white lines) as well as the surrounding free-space. The color-coded force-density plots of $F_y$ (second row) and $F_z$ (third row) also show the vector field ($F_y$, $F_z$) as superposed arrows. In the *p*-polarized case the color scale has been adjusted to exclude high force values (white) at the edges/corners of the half-slab.

Figure 12 shows the *z*-integrated force components, $\int F_y(y, z)\,dz$ and $\int F_z(y, z)\,dz$, plotted versus the *y*-coordinate, for the half-slab depicted in Fig. 11. For *p*-light the net $F_y$ is negative, thus tending to pull the slab to the left. Also $F_z$ is seen to be negative everywhere, indicating that radiation pressure pushes the slab downward. For *s*-light the integrated $F_y$ and $F_z$ show oscillatory behavior between positive and negative values. Overall, however, the negative values dominate, tending to pull the slab to the left and push it downward. Setting the incident *E*-field amplitude to $E_o = 1000$ V/m, the total integrated force ($\iint F_y\,dydz$, $\iint F_z\,dydz$) on the half-slab is found to be $(-0.407, -1.439)$ pN/m for *p*-light and $(-0.313, -1.745)$ pN/m for *s*-light.

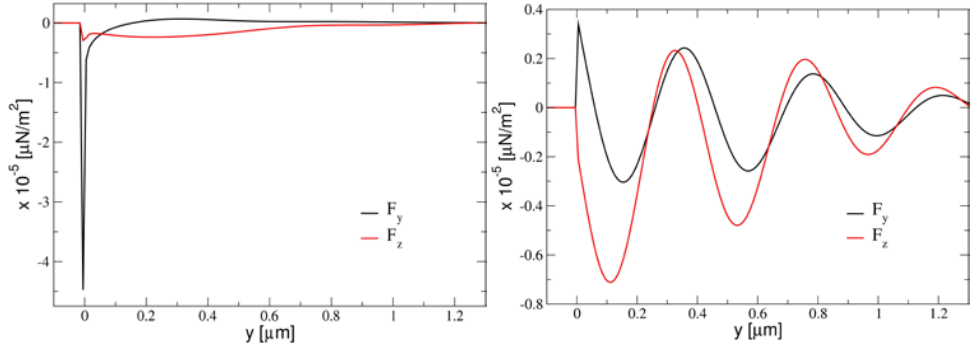

Fig. 12. Force density integrals, $\int F_y(y,z)\,dz$ (black) and $\int F_z(y,z)\,dz$ (red), for the half-slab of Fig. 11 over a range of the *z*-axis that includes the top and bottom facets of the slab. The *z*-integrated force is plotted versus the *y*-coordinate over the entire illuminated region including the slab's vertical side-wall located at *y* = 0. The incident beam is *p*-polarized on the left- and *s*-polarized on the right-hand-side.

## 8. Concluding remarks

We have shown the feasibility of direct computations of the electromagnetic force on dielectric particles based on the Lorentz law of force. The method relies on a numerical computation of the field distributions in and around the object of interest. Once the *E*- and *H*-fields are known, it is straightforward to compute the bound charge density $\rho_b$ and the bound current density $J_b$, then apply the Lorentz law to these charges and currents in order to determine the local distribution of force throughout the volume of interest.

When the particle is surrounded by free space, there are no ambiguities as to the location of the surface charges and the exact value of the force experienced by these charges. This has been the case in the majority of the examples used in the present paper. However, when the particle is immersed in a liquid, the separation of force between the charges that reside on the solid side versus those on the liquid side of the interface becomes non-trivial. In Section 6 we used the average *E*-field across the solid-liquid interface to compute the electric component of the Lorentz force on the (bound) charges induced on the solid side of the interface. The merits and demerits of this particular approach to force computation were discussed in [8], where we also argued that alternative allocations of the radiation force to the solid and liquid bodies are possible, and that the use of these alternate methods may be equally justifiable. Further theoretical and experimental investigations are needed to determine the role of the host liquid as well as the influence of hydrostatic and hydrodynamic forces in trapping experiments.

## Acknowledgments

The authors are grateful to Ewan Wright and Pavel Polynkin for illuminating discussions. This work has been supported by the AFOSR contract F49620-02-1-0380 with the Joint Technology Office, by the *National Science Foundation* STC Program under agreement DMR-0120967, and by the *Office of Naval Research* MURI grant No. N00014-03-1-0793.